___

# Selection of entropy based features for the analysis of the Archimedes' spiral applied to essential tremor

K. López de Ipiña, M. Iturrate, P. Calvo, B. Beitia, J. Garcia-Melero
Universidad del País Vasco/Euskal Herriko Unibertsitatea
{karmele.ipina, mikel.iturrate, itziar.gurruchaga, mariablanca.beitia, joseba.garcia}@ehu.eus

A. Bergareche, P. De la Riva, J.F. Marti-Masso
BioDonostia Health Institute, Donostia, Spain
{jesusalberto.bergarecheyarza, patricia.delarivajuez, josefelix.martimasso}@osakidetza.eus

M. Faundez-Zanuy, E. Sesa-Nogueras, J.Roure
Escola Universitaria Politècnica de Mataró (UPF), Tecnocampus
{faundez, sesa, roure }@tecnocampus.cat

J. Solé-Casals
Data and Signal Processing Group. University of Vic – Central University of Catalonia
jordi.sole@uvic.cat

*Abstract*—: **Biomedical systems are regulated by interacting mechanisms that operate across multiple spatial and temporal scales and produce biosignals with linear and non-linear information inside. In this sense entropy could provide a useful measure about disorder in the system, lack of information in time-series and/or irregularity of the signals. Essential tremor (ET) is the most common movement disorder, being 20 times more common than Parkinson's disease, and 50-70% of this disease cases are estimated to be genetic in origin. Archimedes spiral drawing is one of the most used standard tests for clinical diagnosis. This work, on selection of nonlinear biomarkers from drawings and handwriting, is part of a wide-ranging cross study for the diagnosis of essential tremor in BioDonostia Health Institute. Several entropy algorithms are used to generate non-linear feayures. The automatic analysis system consists of several Machine Learning paradigms.**

## I. INTRODUCTION

Biomedical systems are regulated by interacting mechanisms that operate across multiple spatial and temporal scales and produce biosignals with linear and non-linear information inside. Output variables of real systems often have complex fluctuations that are not only due to noise but also contain information about the intrinsic dynamics and the underlying system. In all cases the dynamics' global aspects can be somehow captured by classic linear methods, but the different approaches are not equivalent to discern all the relevant physical details [1,2]. In this sense the measurement of non-linear features such as the system entropy are essential and useful tools to analyse the system stage. The analysis of system entropy provides not only the probability distributions of the possible state of a system but also the information encoded in it [1]. However the applicability of entropy based methodologies depends on particular characteristics of the data, such as stationarity, time series length, variation of the parameters, level of noise contamination, etc., and important information may be codified also in the temporal dynamics, an aspect which is not usually taken into account [1,3]. Time series generated by biological and biomedical systems most likely contain deterministic and stochastic components [4]. Classical methods of signal and noise analysis can quantify the degree of regularity of a time series by evaluating the appearance of repetitive patterns, but most such methods only model linear components without introducing any information about non-linearity, irregularities or stochastic components. This complex information could be essential when subtle changes are analysed. Massimiliano Zanin et al [1] present a review based on biomedical applications which includes analysis about EEG, anesthesia, cognitive neuroscience or heart rhythms. Among biomedical applications, the related to neurological diseases are a challenge due to their variability and impact in society. Essential tremor is one of the most common.

Essential tremor is a condition that affects individuals worldwide, being 20 times more common than Parkinson's disease. The prevalence of essential tremor (ET) in the western world is of about 0.3-4.0%, 40 years of old males and females are affected more or less equally with an incidence of 23.7 per 100,000 people per year. Studies in the elderly suggest that prevalence in these patients ranges between 3.9% and 14.0%. 50-70% of essential tremor cases are estimated to be genetic in origin [5]. Essential tremor presents itself as a rhythmic tremor (4–12 Hz) that occurs only when the affected muscle is exerting effort. The amplitude of the tremor increases its variability with regard to age but there is no gender predilection. Physical or mental stress could make the tremor worse and the prevalence of Parkinson's disease, in people with essential tremor is greater than in the general population. Parkinson's disease and parkinsonism can also occur simultaneously with essential tremor. With regard to symptoms hand tremor predominates (as it does in Parkinson's disease), and occurs in nearly all cases, followed by head tremor, voice tremor, neck, face, leg, tongue and trunk tremor.

ET is characterized by postural and kinetic tremor which often maximally affects the hands. PD and ET can appear in individuals of the same family [5].

The clinical hallmark and earliest manifestation of the disorder is essential to manage and palliate the symptoms. All these symptoms lead to impaired performance in everyday activities. Approaches to the early diagnosis of ET have in the past few years made significant advances in the development of reliable clinical biomarkers. Despite the usefulness of biomarkers, the cost and technology requirements involved make it impossible to apply such tests to all patients with motor troubles. Given these problems, non-invasive intelligent techniques of diagnosis may become valuable tools for early detection of disorders. Non-technical staff in the habitual environments of the patient could use these methodologies, without altering or blocking the patients' abilities, as speech analysis, handwriting or drawing analysis involved in these techniques is not perceived as a stressful test by the patient. Moreover, these techniques are very low-cost and do not require extensive infrastructure or the availability of medical equipment. They are thus capable of yielding information easily, quickly, and inexpensively [6-8]. It is well established that handwritten tasks can be used for diagnosis of essential tremor. In this sense Archimedes's spiral is one of the most used standard tests in clinical diagnosis [14].

In the past, the analysis of handwriting had to be performed in an offline manner. Only the writing itself (strokes on a paper) were available for analysis. Nowadays, modern capturing devices, such as digitizing tablets and pens (with or without ink) can gather data without losing its temporal dimension. When spatiotemporal information is available, its analysis is referred as online. Modern digitizing tablets not only gather the *x* and *y* coordinates that describe the movement of the writing device as it changes its position, but it can also collect other data, such as the pressure exerted by the writing device on the writing surface, to the azimuth, the angle of the pen in the horizontal plane, the altitude, the angle of the pen with respect the vertical axis [9]. This gives the possibility to analyze not only static ("off-line".) but also dynamic ("on-line") features [10].

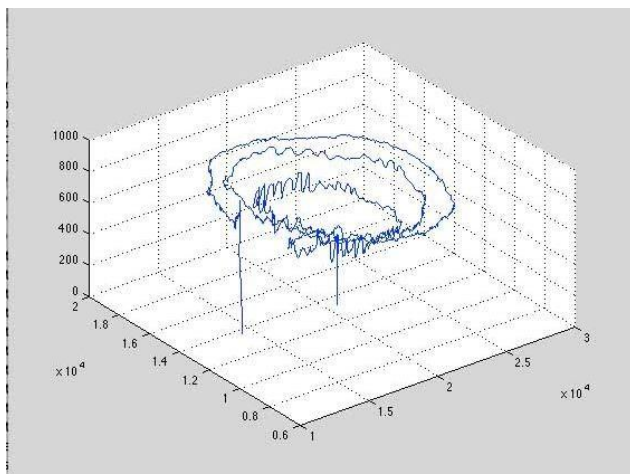

**Figure 1.** . The Archimedes' spiral drawing performed by an individual with essential tremor.

This work is part of a wide-ranging cross study for the diagnosis of essential tremor. The general transversal study is focused to characterize ET (Biodonostia Health Institute) in a study based on families with identified genetics loci. Archimedes's spiral has been selected for the evaluation of nonlinear biomarkers from drawings and handwriting. The presence of integrated features of other diseases such as stress is also analyzed. In the next sections not only classical linear features static and dynamics but also non-linear features based on several entropy algorithms will be analyzed. In that biomarker selection, automatic methodologies will be used. Finally an automatic analysis system based on Machine Learning paradigms measures the quality of the selected features.

## II. MATERIALS

The acquisition is carried out using an Intuos Wacom 4 digitizing tablet. The pen tablet USB [11] captures the following information. The tablet acquired 100 samples per second including the spatial coordinates ($x$, $y$), the pressure, and azimuth and altitude angles. Using this set of dynamic data, further information can be inferred, such as acceleration, velocity, instantaneous trajectory angle, instantaneous displacement, tangential acceleration, curvature radius, centripetal acceleration, etc [12]. The database BIODARW consists of 21 control people (CR) and 29 ET people with identified genetics loci and register of electrophysiological test (EPT) and fMRI. The test consists of a line, the Archimedes' spiral drawing and handwriting with dominant hand and non-dominant hand. Table 1 summarizes the features of the group with ET with regard to EPT, diagnosis and demography [13]. In this work Archimedes' spiral is used (Figure 1). The database presents variability with regard to: tremor frequency, amplitude and pattern, diagnosis scale and demography data (age and gender). From the original database, a subset of the samples of Archimedes's spiral is selected. Moreover from the group with ET, only the samples of the hand with essential tremor are selected. Thus this sub-database BIODARWO consists of 51 samples: 27 samples of the control group and 24 samples of the group with ET.

## III. METHODS

### A. Feature extraction

The research presented here is in the nature of a preliminary experiment; its aim is to define thresholds for a number of biomarkers related to handwriting. It forms part of a broader study focused on early ET detection. Feature search in this work aims at preclinical evaluation so as to formulate useful tests for ET diagnosis [5,13,14].

*1) Linear features*

In this study, we aim at automatically distinguishing of handwriting between an ET patient and a healthy subject by analyzing different linear features (LF) and their variants (max, min, mean and median) in handwriting: time, spatial components, pressure, speed, acceleration, zero crossing rate, and spectral components for on-surface and in-air signals.

*2) Non-linear feature: entropy*

Entropy is a measure of disorder in physical systems, and also a basic quantity with multiple field-specific interpretations. It has been associated with disorder, state-space volume, or lack of information [1,2,15,16]. When try to analyze information content, the Shannon entropy is often considered as the classic, foundational and most natural one [3,4,17]. Richman et al., analyze that entropy, as it relates to dynamical systems, is the rate of information production [18]. On the one hand some authors points that the calculation of entropy usually requires very long data sets that can be difficult or impossible to obtain mainly for biomedical signal. On the other hand methods for estimation of the entropy of a system represented by a time series are not well suited to analysis of the short and noisy data sets encountered in biomedical studies [1,18]. Several proposals for calculating entropy used in this work are presented.

The entropy $H(X)$ of a single discrete random variable $X$ is a measure of its average uncertainty. Shannon entropy [17] is calculated by the equation:

$$H(X) = - \sum_{x_i \in \Theta} p(x_i) \log p(x_i) = - E[\log p(x_i)] \quad (1)$$

Where X represents a random variable with a set of values $\Theta$ and probability mass function $p(x_i) = P_r\{X = x_i\}$, $x_i \in \Theta$, and E represents the expectation operator. Note that $p \log p = 0$ if $p = 0$.

The Approximate Entropy (ApEn) is a statistical measure that smooth transient interference and can suppress the influence of noise by properly setting of the algorithms parameters. It can be employed in the analysis of both stochastic and deterministic signals [19,20]. This is crucial in the case of biological signals, which are outputs of complex biological networks and may be deterministic or stochastic, or both. ApEn provides a model-independent measure of the irregularity of the signals. The algorithm summarizes a time series into a non-negative number, with higher values representing more irregular systems [19,20]. The method examines time series for similar epochs [21]: more frequent and more similar epochs lead to lower values of ApEn. ApEn(m, r, n) measures for n points, sequences that are similar for m points remain similar, within a tolerance r, at the next point. Thus a low value of ApEn reflects a high degree of regularity. ApEn algorithm counts each sequence as matching itself. In order to reduce this bias, Sample Entropy SampEn(m, r, n) was developed, which does not count self-matches.

The sample entropy statistic $SampEn(m, r, n)$ is defined as:
$$SampEn(m, r, n) = \lim_{n \to \infty} \{-\ln(A^m(r)/B^m(r))\}$$
$$= -\ln(A/B), \quad (2)$$
with
$$A = [(n - m - 1)(n - m)/2]A^m(r), \quad (3)$$
and
$$B = [(n - m - 1)(n - m)/2]B^m(r). \quad (4)$$

$B^m(r)$ is the probability that two sequences match for $m$ points. Similarly, $A^m(r)$ is the probability that two sequences match for $m + 1$ points:

$$A^m(r) = (n - m)^{-1} \sum_{i=1}^{n-m} A_i^m(r), \quad (5)$$

The scalar $r$ is the tolerance for accepting matches. In the present investigation, we used the parameters recommended in [22], with $m = 2$ and $r = 0.2$ (standard deviation of the sources is normalized to 1). SampEn is a robust quantifier of complexity as for instance EEG signals [23], and can be used to detect of artifacts in EEG recordings [24].

Permutation entropy directly accounts for the temporal information contained in the time series; furthermore, it has the quality of simplicity, robustness and very low computational cost [1,3,4]. Bandt and Pompe [25] introduce a simple and robust method based on the Shannon entropy measurement that takes into account time causality by comparing neighboring values in a time series. The appropriate symbol sequence arises naturally from the time series, with no prior knowledge assumed [1]. The Permutation entropy is calculated for a given time series $\{x_1, x_2, ..., x_n\}$ as a function of the scale factor $s$. In order to be able to compute the permutation of a new time vector $X_j$, $S_t = [X_t, X_{t+1}, ..., X_{t+m-1}]$ is generated with the embedding dimension $m$ and then arranged in an increasing order: $[X_{t+j_1-1} \le X_{t+j_2-1} \le \cdots \le X_{t+j_n-1}]$. Given $m$ different values, there will be m! possible patterns $\pi$, also known as permutations. If $f(\pi)$ denotes its frequency in the time series, its relative frequency is $p(\pi) = f(\pi)/(L/s - m + 1)$. The permutation entropy is then defined as:

$$PE = -\sum_{i=1}^{m!} p(\pi_i) \ln p(\pi_i) \quad (6)$$

*B. Automatic classification*

The main goal of the present work is feature search in handwriting aiming at preclinical evaluation in order to define tests for ET diagnosis. These features will define the control group (CR) and the essential tremor group (ET). A secondary goal is the optimization of computational cost with the aim of making these techniques useful for real-time applications in real environments. Thus, automatic classification will be modeled with this in mind. We used five different classifiers:

- A Support Vector Machine (SVM) with polynomial kernel
- A multi layer perceptron (MLP) with neuron number in hidden layer (NNHL) = max(Attribute/Number+Classes/Number) and training step (TS) = NNHL*10
- k-NN Algorithm.

The WEKA software suite [26] has been used to carry out the experiments. The results were evaluated using Classification Error Rate (CER, %). For training and validation steps we used k-fold cross-validation with k=10. Cross-validation is a robust validation method for variable selection [27]. Repeated cross-validation (as calculated by the WEKA environment) allows robust statistical tests. We also use the measurement provided automatically by WEKA "Coverage of cases" (0.95 level) that is the confidence interval at 95% level.

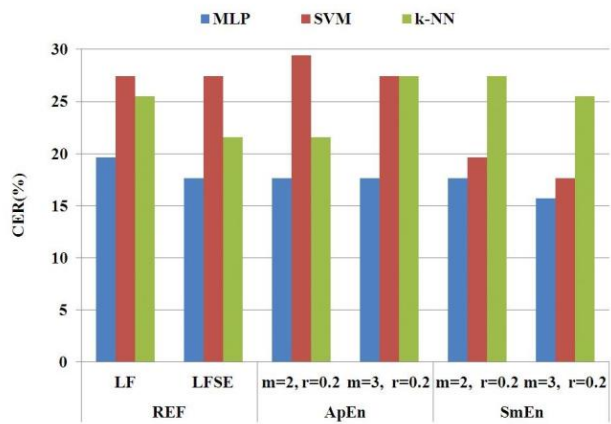

**Figure 2.** CER (%) for paradigms with linear and non-linear features sets based on entropy: Shannon, ApEn, SmEn and Permutation Entropy.

## IV. RESULTS AND DISCUSSION

The experimentation has been carried out with the balanced subset BIODARWO. The goal of these experiments was to examine the potential of entropy algorithms and selected features for automatic measurement of the degradation of Archimedes's spiral drawing with ET. Thus, previously defined feature sets have been evaluated in order to properly define control and ET groups. In a first stage linear and non-linear features have been extracted by several methods described in section III: Shannon entropy, ApEn and SmEn. Automatic classification by described (section III) paradigms was performed over the database. The results of CER (%) for paradigms with linear and non-linear features sets are summarized in Figure 2. For both algorithms m=2,3 and tolerance r=0.2.

- Non Linear Features (NLF) sets increase about 7% the features number
- Non-linear features sets improve the results for all the paradigms
- Shannon entropy outperforms LF (LFSE) for MLP and k-NN
- ApEn improve system performance for m=3 with MLP and SVM. k-NN has better performance for m=2.
- SmEn with m=3 appears as the best option for all paradigms.
- The best option is SmEn with m=3 and MLP with a CER of 15.69%.

In a second stage permutation entropy is evaluated for different orders m and time delay. In our particular case and due the signals are composed from 5000-1000 samples, and m parameter was fixed until m=7. The results are also compared with SmEn with m=3 and tolerance of r=0.2. The results are shown in Figure 3.

- PE improves the results in most of the cases
- The best results are obtained with PE and m=7, t=7.
- MLP obtains the best results for this last option
- The best option is PE-m7t7 with MLP and CER of 15.65%.
- Good results are achieved even with k-NN with less computational cost.

## V. CONCLUSIONS

This work, on selection of nonlinear biomarkers from drawings and handwriting, is part of a wide-ranging cross study for the diagnosis of essential tremor which is developed in Biodonostia Health Institute. Specifically the main goal of the present work is the analysis of features in Archimedes's spiral drawing, one of the most used standard tests for clinical diagnosis of ET. In this sense entropy based features have been add to a set of classical linear features (static and dynamics). Several entropy algorithms have been evaluated by an automatic analysis system consists of several Machine Learning. The best option is MLP with permutation entropy and good results are obtained even with k-NN. Then these new biomarkers will be integrated in future works with those obtained in the Biodonostia study. It should be highlighted that the use of this technology could provide undoubted benefits towards the development of more sustainable, low cost, high quality, non-invasive technologies. These systems are easily adaptable to the user and environment, and from a social and economic point of view can be very useful in real complex environments. In future works new non-linear features, entropy algorithms and automatic selection methodologies will be used.

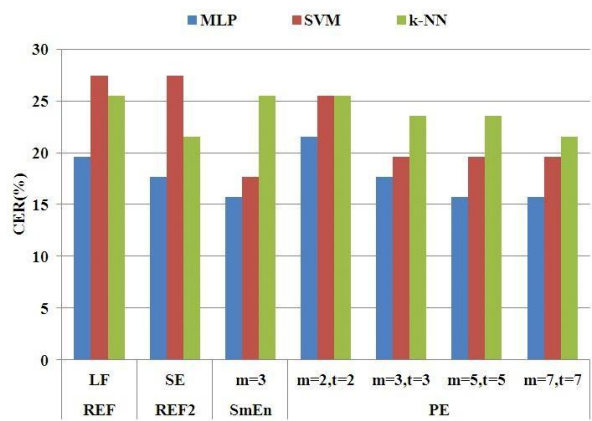

**Figure 3.** CER (%) for the paradigms with the references Linear Features (LF) and Shannon Entropy (SE) and other entropy paradigms

___

*Acknowledgments*

This work has been partially supported by the University of the Basque Country by UPV/EHU—58/14 project, SAIOTEK from the Basque Government, University of Vic-Central University of Catalonia under the research grant R0904, and the Spanish Ministerio de Ciencia e Innovación TEC2012-38630-C04-03.